\documentstyle[12pt]{article}
\begin{document}
\begin{center}
{\Large{\bf Hidden Symmetries in Second Class Constrained Systems - 
Are New Fields Necessary?}}

\vspace{12mm}
{\large{\em A S Vytheeswaran}}{\footnote{\bf email : 
buniphy@kar.nic.in; ~~asmdevan@bgl.vsnl.net.in; ~~vythee@rediffmail.com}}

\vspace{5mm}
{{\sf Department of Physics, \\
Bangalore University, Jnana Bharathi Campus,\\
Bangalore 560 056, ~INDIA}}
\end{center}

\begin{abstract}
For many systems with second class constraints, the question posed in 
the title is answered in the negative. We prove this for a range of 
systems with two second class constraints. After looking at two examples,
we consider a fairly general proof. It is shown that, to unravel gauge 
invariances in second class constrained systems, it is sufficient to 
work in the original phase space itself. Extension of the phase space 
by introducing new variables or fields is {\em not required}.
\end{abstract}

\vspace{1.2cm}



\newpage

\section{{Introduction}}

The conversion of systems with second class constraints into those with
first class ones has been of interest in recent times. Since first class 
constraints are generators of gauge transformations such conversions are  
useful in having a better and more illuminating view of second class 
constrained systems. The unearthing of inherent gauge symmetries, implied 
by the modification into first class constraints, allows a broader study 
of the system, in contrast to the limited view offered by the original 
second class constraints. 

The basic premise behind such a conversion is that the second class 
constrained system is considered to be a gauge fixed version of a gauge 
theory; the latter goes back to the former under a certain set of gauge 
fixing conditions. The advantage in having a gauge theory lies in the 
fact that other gauges can also be considered, sometimes more profitably. 
Further such conversions into gauge theories can result in more than one 
gauge theory for the same second class system, with some gauge theories 
being more relevant than the others. This also raises the interesting 
possibility of (many) inequivalent gauge - fixed versions for the same 
gauge theory. 

The motivation for this conversion into gauge theories came originally 
from anomalous gauge theories. In these theories the classical 
gauge invariance is lost upon quantisation. In terms of constraints, this 
means the classical first class constraints become second class upon 
quantisation. In this context, the conversion to gauge theories would 
mean recovering the lost gauge invariance. 

There are basically two ideas proposed to convert second class constrained 
systems into gauge theories. One idea, proposed by Faddeev and Shatashvili 
\cite{FaSha}, uses an {\em enlarged} phase space; the other, given in 
\cite{RaPa}, is confined to the {\em original} phase space itself. Both 
ideas are based on the possibility that a system with second class 
constraints can be considered to be a gauge - fixed version of some
gauge theory. 

Based on these two ideas, methods have been developed and applied to 
realise hidden symmetries in various systems. While the Batalin - Fradkin 
method \cite{BaFa} follows the Faddeev - Shatashvili idea, the Gauge 
Unfixing method of \cite{RaPa,RaVy} uses the {\em original} phase space.
There are other related methods too; while the one given by Wotzasek 
\cite{Wot} uses an extended phase space, the method of Bizdadea and
Saliu \cite{BizSa} is developed in the original phase space, with BRST 
quantisation.

Even though the Batalin - Fradkin and the Gauge Unfixing formulations 
appear quite different, when applied to various systems they give 
essentially the same results!  The first class constraints may look 
different, but relevant observables obtained by demanding their gauge 
invariance in both the methods are essentially the same. In this context 
we refer to \cite{Asvva}, which compares results of the two methods 
applied to the chiral Schwinger model, the Proca 
model and abelian Chern-Simons theory. For these theories it was found 
that, in both classical and path integral context, the gauge 
invariant Hamiltonians and the actions obtained using both methods are 
the same! Hence, as far as these systems were concerned, an enlarged 
phase space was found to be {\em not really necessary} to obtain the 
hidden gauge invariances.

In this paper, we pursue this matter further and compare the two methods 
in a more general context. As a first step towards demonstrating this 
formal equivalence, we consider two examples and apply and compare the 
two methods. For the general case, to simplify matters, we consider 
only two second class constraints. We will see that even in a fairly 
general context, the two methods when compared on their respective 
first class constrained surfaces give {\em equivalent results}. Hence
we show that, contrary to widely accepted belief, {\em extra fields are
not really necessary} for inducing gauge symmetries in second class 
constrained systems.

In Section 2 we review the two methods for the case of two second class 
constraints. In Section 3 we look at two specific systems, the chiral 
Schwinger model and the non-linear sigma model. In Section 4 we present a 
fairly general proof, and conclude in Section 5.

\section{The Formalisms}

We consider a finite dimensional system \cite{Dirac} with phase space 
co-ordinates $ q^i $ and conjugate momenta $ p_i ~(i = 1,2,\ldots  N). 
$ The system has two second class constraints,
\begin{equation}
Q_1(q,p) \approx 0, \hspace{1.4in} Q_2(q,p) \approx 0,
\end{equation}
defining a constraint surface $ \sum_2$. Due to their second class nature, 
the $2 \times 2$ antisymmetric matrix ${\cal E}$ whose elements ${\cal 
E}_{ab}$ are Poisson brackets among the $ Q$'s,
\begin{equation}
{\cal E}_{ab}(q,p) = \{ Q_a, Q_b \}\hspace{1in}a,b=1,2,
\end{equation}
is {\em invertible} everywhere, even on the surface $ \sum_2. $ The 
canonical Hamiltonian is $ H_c $ and the total Hamiltonian is 
\begin{equation}
H = H_c + \mu_1Q_1 + \mu_2Q_2,
\end{equation} 
where the multipliers $ \mu_1, \mu_2  $ are determined by demanding the
consistency conditions $ \{ Q_a, H \} = 0, ~a = 1,2 $ on the surface $ 
\sum_2 $. Other relevant physical quantities must also have similar 
properties with respect to the $ Q_a. $ These considerations can also 
be extended to field theories.

\subsection{{Batalin - Fradkin (BF) method}}

As mentioned in the Introduction, this method \cite{BaFa} is formulated 
in an enlarged phase space, the extent of enlargement depending on the 
number of second class constraints. Here since this number is two, the 
phase space is enlarged by introducing two new variables $ \Phi^a 
(a=1,2) $. The enlarged phase space $ (q,p,\Phi) $ has the basic Poisson 
brackets
\begin{equation}
\{ q^i, p_j \} = \delta^i_{\;j}, \hspace{1.6in} \{\Phi^a,\Phi^b\} = 
\omega^{ab},
\end{equation}
with all other Poisson brackets zero. The antisymmetric $ 2\times 2 $ 
matrix $ \omega^{ab} $ is a constant matrix, unspecified for the present.

The first class constraints are obtained as functions in this {\em 
extended} phase space. Since we had initially two second class constraints,
 there will now be {\em two} first class constraints, given in general by
\begin{eqnarray}
{\widetilde Q_a}(q^i,p_i,\Phi^a) & = & Q_a + \sum_{m = 1}^{\infty} 
Q_a^{(m)}, \hspace{1.2in} Q_a^{(m)} \sim (\Phi^a)^{m} \\
{\widetilde Q_a}(q^i,p_i,0) & = & Q_a \nonumber
\end{eqnarray}
where the second line gives the boundary condition. The terms of various 
orders in the expansion for $ {\widetilde Q_a} $ are obtained by 
demanding that the $ {\widetilde Q_a} $ are strongly first class,
\begin{equation}
\{ {\widetilde Q_a}, {\widetilde Q_b} \} = 0, \hspace{1in}a,b = 1,2.
\end{equation}
For instance for the lowest order this requirement gives  
\begin{equation}
{\cal E}_{ab} = - X_{ac}(q,p)\omega^{cd}X_{db},
\end{equation}
where the matrix $ X $ is also unspecified for the present. 
Using (2) and (4), eqn. (7) can be satisfied, if we write and substitute
\begin{equation}
Q_a^{(1)} = X_{ab}(q,p)\Phi^b,
\end{equation}
in (5) and consider terms at lowest order in (6). Taking $ \omega_{ab} 
$ and $ X^{ab} $ to be inverses to $ \omega^{ab} $ and $ X_{ab} $ 
respectively, the higher order terms are given by 
\begin{eqnarray}
Q_a^{(n+1)} & = & -\frac{1}{(n+2)} \;\Phi^b\omega_{bc}X^{cd}B^{(n)}_{da},
\hspace{11mm} n \geq 1, \nonumber\\
B_{ab}^{(1)} & = & \rule{0mm}{8mm} \{ Q_{[a}, Q_{b]}^{(1)} \}_{(q,p)}\\
B_{ab}^{(n)} = \frac{1}{2}B^{(n)}_{[ab]} & = & \rule{0mm}{8mm}\sum_{m=0}^n 
\{ Q_a^{(n-m)}, Q_b^{(m)} \}_{(p,q)} ~+ ~\sum_{m=0}^{n-2} \{ Q_a^{(n-m)}, 
Q_b^{(m+2)} \}_{(\Phi)} \hspace{1.1cm} n \geq 2,\nonumber
\end{eqnarray}
where the square brackets in the subscript imply antisymmetrization. 
In the last two lines in (9) the subscript $(q,p)$ implies evaluation 
of the corresponding Poisson bracket with respect to only the $ (q^i,
p_i) $, while the subscript $ (\Phi) $ implies evaluation with respect 
to only the $\Phi$. Further, in the above equations the matrix $X$ 
along with the matrix $ \omega $ (and hence the $ \Phi^a $) are chosen 
according to convenience. This implies an inherent {\em arbitrariness} 
in our choice of a convenient gauge theory.

It is important to note that eqn. (7) can always be written so for the 
case of 2 constraints. For more than two second class constraints, this 
has to be taken as an assumption, which however may not hold in a very 
general context. In a sense, the matrix $ X $ can be called the 
``square root'' of the matrix ${\cal E}$. We will come back to this 
issue later.

To get gauge invariant observables, we note that in general relevant 
quantities of the original second class system cannot be used here 
directly, since they are not invariant (i.e., do not have zero Poisson 
brackets) with respect to the new first class constraints. They are 
made gauge invariant by modifying them in the extended phase space. 
For a function $A(q,p)$ on the original phase space, the corresponding 
gauge invariant variables are,
\begin{equation}
{\tilde A}(q^i,p_i,\Phi) = A + \sum_{m=1}^{\infty} A^{(m)} \hspace{1.1in} 
A^{(m)} 
\sim (\Phi^a)^{m},
\end{equation}
with the terms of various orders obtained by demanding that
\begin{equation}
\{ {\tilde A}, {\widetilde Q_a} \} = 0 \hspace{1in} a = 1,2.
\end{equation}
The terms $ A^{(m)}$ in the expansion (10) are
\begin{eqnarray}
A^{(m+1)} & = & -\frac{\displaystyle 1}{\displaystyle (m+1)}\Phi^a
\omega_{ab}X^{bc}G^{(m)}_c \hspace{1.5in} m \geq 0,\nonumber\\
G^{(0)}_a & = & \rule{0mm}{8mm}\{ Q_a, A \} \\
G_c^{(1)} & = & \rule{0mm}{8mm}\{ Q_c^{(1)}, A \} + \{Q_c, A^{(1)} \} + 
\{ Q_c^{(2)}, A^{(1)} \}_{(\Phi)}\nonumber\\ 
G_c^{(m)} & = & \rule{0mm}{8mm}\sum^{m}_{n=0} \{ Q_c^{(m-n)}, 
A^{(n)} \}_{(q,p)} + \sum_{n=0}^{m-2} \{ Q_c^{(m-n)}, A^{(n+2)} 
\}_{(\Phi)} + \{ Q_c^{(m+1)}, A^{(1)} \}_{(\Phi)} ~~~~m\geq 2,\nonumber
\end{eqnarray}
where in the last line, the subscripts $ (q, p)$ and $ (\Phi) $ stand for
 evaluation of corresponding Poisson brackets with respect to ($q^i, p_i$)
 and $ \Phi $ respectively. Thus in this method, the first class 
constraints $ {\widetilde Q}_a \approx 0 $ and the various gauge invariant
observables $ {\widetilde A} $ describe the new gauge theory. 

\subsection{{The Gauge Unfixing (GU) method}}

This method \cite{RaVy}, in stark contrast to the BF method, makes no 
enlargement of the phase space while extracting a gauge theory from a 
second class constrained system. Rather, since the number of second class 
constraints is even (we consider here only bosonic constraints), this 
method attempts to treat half these constraints to form a first 
class subset, and the other half as the corresponding gauge fixing subset. 
This latter subset is discarded, retaining only the first class subset, 
and so we have a gauge theory. 

In a general system, getting a first class subset is a non-trivial issue 
\cite{RaVy}; this might be possible only under certain conditions. However 
in the case of only two second class constraints, the first class 
constraint can {\em always} be chosen.

For instance, we can choose $ Q_1 $ as our first class constraint, and $ 
Q_2 $ as its gauge fixing constraint. We redefine, using (2),
\begin{equation}
Q_1 \rightarrow \chi = {\cal E}_{12}^{-1}Q_1, \hspace{1.4in} Q_2 \rightarrow 
\psi,
\end{equation}
and discard the $ \psi $ as a constraint (i.e., no longer consider $ 
\psi = 0 $). To obtain the gauge invariant 
Hamiltonian and other physical quantities we construct a projection 
operator $ I\!\!P $ by defining its operation on any phase space function 
$A$ as 
\begin{equation}
I\!\!{ P}(A) = {\widetilde A} \equiv ~: e^{-\psi{\hat \chi}} : A ~= 
A - \psi \{\chi, A \} + \frac{1}{2!}\psi^2 \{ \chi, \{ \chi, A \}\}
- \frac{1}{3!} \psi^3 \{\chi, \{ \chi, \{ \chi, A \}\}\} + \ldots - 
\ldots
\end{equation}
where it may noted that the $ \psi $ is always outside the Poisson 
brackets on the right hand side. The gauge invariant quantities are the
$ I\!\!P(A) = {\widetilde A} $, since they satisfy the gauge invariance
condition $ \{ \chi, {\widetilde A} \} = 0 $. These and the first class
constraint $ \chi = 0 $ describe the new gauge theory.

It must be noted that even in this method, there is an inherent {\em 
arbitrariness}; of the two second class constraints the first class 
constraint can be chosen in two ways. The two choices define two 
different projection operators, and the gauge theories so constructed 
will in general be different. This arbitrariness can be exploited to 
advantage. 

\section{{Examples}}

\subsection{{The chiral Schwinger model}}

This well known anomalous gauge theory \cite{JaRa} involves chiral 
fermions coupled to a $U(1) $ gauge field in $(1+1)$ dimensions. 
Classically the theory 
has gauge invariance, but this is lost upon quantisation. We look at 
its bosonised version, the advantage being that the corresponding 
classical theory itself has no gauge invariance. We have
\begin{equation}
{\cal L} = -\frac{1}{4}F_{\mu\nu}F^{\mu\nu} + 
\frac{1}{2}(\partial_{\mu}\phi)^2 + e(g^{\mu\nu} - 
\epsilon^{\mu\nu})(\partial_{\mu}\phi)A_{\nu} + \frac{1}{2}e^2\alpha 
A_{\mu}^2,
\end{equation}
where $g^{\mu\nu} $ = diag$(1,-1)$, ~$ \epsilon^{01} = - ~\epsilon^{10} 
= 1 $ and $ \alpha $ is the regularisation parameter. The Lagrangian 
is gauge non-invariant for all values of $ \alpha. $ We consider the 
case $ \alpha > 1.$

The canonical Hamiltonian density is
\begin{eqnarray}
{\cal H}_c & = & \frac{1}{2}\pi_1^2 + \frac{1}{2}\pi_{\phi}^2 + 
\frac{1}{2}(\partial_{1}\phi)^2 + e(\partial_1\phi + \pi_{\phi})A_1 + 
\frac{1}{2}e^2(\alpha + 1)A^2_{1} \nonumber \\
&& \rule{0mm}{8mm}\hspace{0.6in} - A_0\left [- \partial_1\pi_1 + 
\frac{1}{2}e^2(\alpha - 1)A_0 + e(\partial_1\phi + \pi_{\phi}) + e^2A_1 
\right ]
\end{eqnarray}
where $ ~\pi_1 = F^{01} = \partial^0A^1 - \partial^1A^0 $ and $ 
~\pi_{\phi} = \partial_0\phi + e(A_0 - A_1) $ are the momenta 
conjugate to $ A_1 $ and $\phi$ respectively. The constraints are 
\begin{eqnarray}
Q_1 & = & \pi_0 \approx 0\nonumber \\
Q_2 & = & - ~\partial_1\pi_1 + e^2(\alpha - 1)A_0 + e(\partial_1\phi + 
\pi_{\phi}) + e^2A_1 \approx 0,
\end{eqnarray}
defining a constraint surface $ \sum_2. $ These are of the second class, 
\begin{equation}
{\cal E}_{12} = \left\{ Q_1(x), Q_2(y) \right\} = -e^2(\alpha - 1) 
\delta(x - y).
\end{equation}

\vspace{3mm}
Following the BF method \cite{Kim}, the phase space is extended by 
introducing two 
fields $ \Phi^1, \Phi^2 $, with Poisson bracket relations of the 
form (4). The new first class constraints have the general form (5), with 
the first order term as given in (8). As mentioned earlier, there is a 
natural arbtrariness in choosing the matrices $ \omega^{ab} $ and $ X_{ab}. 
$ The choice
\begin{eqnarray}
\omega = \left( \begin{array}{c}
	0 ~~~ 1\\
       -1 ~~ 0 
\end{array}                   \right) \delta(x-y) \hspace{0.8in}
X (x,y) = e\sqrt{\alpha-1}\left( \begin{array}{c}
				1 ~~ 0\\
				0 ~~ 1 
\end{array}		\right) \delta(x-y)
\end{eqnarray}
allows the two new fields to form a canonically conjugate pair. The 
higher order terms beyond the first in the expansion (5) are all zero. 
Then the first class constraints are
\begin{equation}
{\widetilde Q_a} = Q_a + e\sqrt{\alpha-1}\,\Phi^a, \hspace{1in} a = 1,2,
\end{equation}
which, using (4), (18) and (19), can be verified to be strongly first class.

Using the general expressions in (10) and (12) the gauge invariant 
Hamiltonian for the choice (19) is 
\begin{eqnarray}
{\widetilde H}_{BF} & = & H_c + {\displaystyle \int}dx\left[ 
- ~\frac{\displaystyle \left(e\pi_1 + e(\alpha-1)\partial_1A_1\right)
}{\displaystyle \sqrt{\alpha-1}} ~\Phi^1 + \frac{\displaystyle e^2}{
\displaystyle 
2(\alpha-1)}(\Phi^1)^2 \right.\nonumber \\
&& \rule{0mm}{8mm}\hspace{1.8in}\left.+ ~\frac{1}{2}(\partial_1\Phi^1)^2 ~+ 
\frac{1}{2}(\Phi^2)^2 - \frac{{\widetilde Q_2}\Phi^2}{\displaystyle 
e\sqrt{\alpha-1}} \right],
\end{eqnarray}
with $ H_c $ given by (16). This $ {\widetilde H_{BF}} 
$ has zero PBs with the constraints in (20).

\vspace{4mm}
Coming to the {\em Gauge Unfixing} (GU) method \cite{Asv}, we reiterate that no 
new field need be introduced. The first class constraint is taken to 
be just one of the two existing constraints. We choose, after a rescaling
\begin{equation}
\chi = \frac{1}{\displaystyle e^2(\alpha-1)} ~Q_2 ~,
\end{equation}
so that the relevant constraint surface $ \sum_1 $ is defined by $ \chi 
\cong 0.$ The gauge fixing-like constraint is $ \psi = 0,$ and is discarded 
(that is {\em unfixed} ). The gauge invariant Hamiltonian is obtained by 
constructing a projection operator $ I\!\!{ P} $ of the form (14) and 
using it on the $H_c$. We get $ I\!\!P(H_c) = {\widetilde H}_{GU} $, 
\begin{equation}
{\widetilde H}_{GU} = H_c + {\displaystyle \int}dx\left[
\frac{\displaystyle \left(\pi_1 + (\alpha-1)\partial_1A_1\right)}{
\displaystyle {\alpha-1}} \;Q_1 + \frac{\displaystyle (\partial_1Q_1)^2}{
\displaystyle 2e^2(\alpha - 1)} + \frac{\displaystyle Q_1^2}{\displaystyle 
2(\alpha-1)^2} \right],
\end{equation}
which satisfies  $ \{\chi, {\widetilde H}_{GU} \} = 0.$

\vspace{1mm}
It can be seen that, if we make the identification $ \Phi^1 = - 
\;\frac{\displaystyle Q_1}{\displaystyle e\sqrt{\alpha - 1}} $, the $ 
{\widetilde H}_{BF} $ in (21) and the $ {\widetilde H}_{GU} $ in (23) 
are almost the {\em same}. The difference between these two Hamiltonians 
are the extra terms $ {\displaystyle \int} dx \left (\frac{\displaystyle 
(\Phi^2)^2}{\displaystyle 2} - \frac{\displaystyle \Phi^2}{\displaystyle 
e\sqrt{\alpha - 1}}{\widetilde Q_2} \right), $ appearing in (21). The 
second of these is zero due to (20). The first term, when rewritten 
using (20), is proportional to $ {\widetilde Q}_2 $ and the 
constraint $ \chi $ in (22).

\vspace{1mm}
We emphasise the two rather {\em different paths} used to get these 
Hamiltonians. 
One requires the introduction of an extra (canonical) pair of fields, 
while the other doesn't need this. In both cases extra terms are needed 
to make the original Hamiltonian gauge invariant. For the $ {\widetilde 
H_{BF}} $ these terms had to be written down using the extra fields, 
whereas in the $ {\widetilde H_{GU}} $ these terms involve a variable 
{\em already present} in the original theory.

\vspace{4mm}
We look at the path integral quantisation for these two gauge invariant 
Hamiltonians. For the Batalin-Fradkin Hamiltonian $ {\widetilde H_{BF}}$, 
we first redefine
\begin{equation}
\Phi^1 \rightarrow \theta\hspace{1.2in} \Phi^2 \rightarrow \pi_{\theta},
\nonumber
\end{equation} 
and the partition function is 
\begin{eqnarray}
{\cal Z}_{BF} & = & {\displaystyle \int}{\cal D}(\pi_{\mu}, A^{\mu}, 
\pi_{\phi}, \phi, \theta, \pi_{\theta}, \lambda_1, \lambda_2)~e^{iS_{BF}} \\
S_{BF} & = & {\displaystyle \int} dxdt \left [\pi_0{\dot A^0} + 
\pi_1{\dot A^1} + \pi_{\phi}{\dot \phi} + \pi_{\theta}{\dot \theta} 
- {\widetilde{\cal H}}_{BF} - \lambda_1{\widetilde Q_1} - \lambda_2{\widetilde Q_2} 
\right].
\nonumber
\end{eqnarray}
Here $ \lambda_1, \lambda_2 $ are undetermined Lagrange multipliers 
corresponding to the first class constraints $ {\widetilde Q_1}, 
{\widetilde Q_2} $ respectively. The integration over the $ \pi_0 $
gives the delta function $ \delta({\dot A}_0 - \lambda_1)$, which 
can be used while integrating over the $ \lambda_1 $. We next make the 
transformations 
$$ 
\begin{array}{rcl}
A_0 \rightarrow A_0^{\;\prime} = A_0 - \lambda_2 + \frac{\displaystyle 
\pi_{\theta}}{\displaystyle e\sqrt{\alpha - 1}}, ~~~~~~&& ~~~~\pi_1 
\rightarrow \pi_1^{\;\prime} = \pi_1 + \partial_0A_1 - \partial_1A_0^{
\;\prime} - \frac{\displaystyle e \;\theta}{\displaystyle \sqrt{\alpha 
- 1}},\\
\pi_{\phi} \rightarrow \pi_{\phi}^{\;\prime} = \pi_{\phi} - {\dot \phi} - 
e(A_0^{\;\prime} - A_1),  ~~~~&& \rule{0mm}{8mm}~~~~
\lambda_2 \rightarrow \lambda^{\prime}_2 = e\sqrt{\alpha - 1}\;\lambda_2
- {\dot \theta}
\end{array}
$$
 and after rearranging terms, we get the action to be 
\begin{eqnarray}
S_{BF} & = & {\displaystyle \int} dxdt \left[ - 
\frac{1}{2}(\pi_1^{\prime})^2 - \frac{1}{2}(\pi_{\phi}^{\prime})^2 - 
\frac{1}{2}(\lambda_2^{\prime})^2 + 
\frac{1}{2}(\partial_0A_1 - \partial_1A_0^{\prime})^2 + \frac{1}{2}
(\partial_{\mu}\phi)^2 \right.\nonumber\\
&& \hspace{0.7in} \left. + ~e({\dot \phi}{A_0^{\prime}} - \partial_1A_1) 
- e({\dot \phi}A_1 - A_0^{\prime}\partial_1\phi) + 
\frac{e^2\alpha}{2}[(A_0^{\prime})^2 - A_1^2]  \right.\\
&& \hspace{0.7in}\left. + \frac{1}{2}(\partial_{\mu}\theta)^2 - 
e\;\theta\sqrt{\alpha-1}\;({\dot A_0^{\prime}} - \partial_1A_1) - 
~\frac{\displaystyle e\;\theta}{\displaystyle \sqrt{\alpha - 1}}({\dot 
A_1} - \partial_1A_0^{\prime})  \right].\nonumber
\end{eqnarray}
Putting this in the path integral, the $ \pi_1^{\;\prime}, ~\pi_{\phi}^{\;\prime}, ~\lambda_2^{\;\prime}, 
$ are integrated over. We redefine $ \theta^{\;\prime} = \frac{
\displaystyle \theta}{\displaystyle \sqrt{\alpha - 1}}, $ and dropping 
the primes on $ \theta^{\;\prime} $ and $ A_0^{\prime} $, we get

\begin{eqnarray}
{\cal Z}_{BF} & = & {\displaystyle \int} {\cal D}(A^{\mu}, \phi, \theta) 
~~e^{iS_{BF}} \nonumber \\
S_{BF} & = & {\displaystyle \int} dxdt \left( - \frac{1}{4}F_{\mu\nu}
F^{\mu\nu} + \frac{e^2\alpha}{2}A_{\mu}A^{\mu} + e(\nu^{\mu\nu} - 
\epsilon^{\mu\nu})(\partial_{\mu}\phi)A_{\nu}  \right.\\
&& \hspace{0.8in} \left. + \frac{1}{2}(\partial_{\mu}\phi)^2 + 
\frac{\alpha-1}{2}(\partial_{\mu}\theta)^2 - e\theta\left[(\alpha - 1)
\eta^{\mu\nu} + \epsilon^{\mu\nu}\right](\partial_{\mu}A_{\nu})\right).
\nonumber
\end{eqnarray}
The action $ S_{BF} $ above is just the gauge invariant version of the 
chiral Schwinger model. As is well known, this action was obtained 
earlier by adding the (Wess - Zumino) terms \cite{WeZu} in the variable 
$ \theta $ to the original bosonised action (15). Other arguments have 
also been used to get the same result \cite{FaSha, MoOz}. In the 
Batalin-Fradkin approach, these Wess Zumino terms and $ \theta $ come up 
due to the enlargement of the phase space.

\vspace{4mm}
In the Gauge Unfixing method, the path integral is
\begin{equation}
{\cal Z}_{GU} = {\displaystyle \int} {\cal D}(A^{\mu}, \pi_{\mu}, \phi, 
\pi_{\phi}, \mu) ~exp \left(i{\displaystyle \int} dxdt \left 
[\pi_0{\dot A^0} + \pi_1{\dot A^1} + \pi_{\phi}{\dot \phi} - {\cal 
{\widetilde H}}_{GU} - \mu\chi \right]\right),
\end{equation}
with $ {\widetilde H}_{GU} $ given by (23). Here $ \mu $ is the 
arbitrary Lagrange multiplier. We make the transformations 
$$
\begin{array}{rcl}
 A_0 \rightarrow A_0^{\prime} = A_0 - \frac{\displaystyle \mu
}{\displaystyle e^2(\alpha-1)}, ~~~~~& & ~~~~~\pi_1 \rightarrow 
\pi_1^{\;\prime} = \pi_1 + \partial_0A_1 - \partial_1A_0^{\prime} + 
\frac{\displaystyle \pi_0}{\displaystyle \alpha - 1},\\
\pi_{\phi} \rightarrow \pi_{\phi}^{\;\prime} = \pi_{\phi} - {\dot \phi} 
+ eA_1 - eA_0^{\prime}, && ~~~~~\mu \rightarrow ~\mu^{\prime} 
\;= \mu + \partial_0\pi_0.
\end{array}
$$ 
Dropping the prime on the $A_0^{\prime}$  and integrating over $ 
\pi_1^{\;\prime}, \pi_{\phi}^{\;\prime}$ and $ \mu^{\;\prime} $ we get
\begin{eqnarray}
{\cal Z}_{GU} & = & {\displaystyle \int} {\cal D}(A^{\mu}, \phi, \pi_0) 
~e^{iS_{GU}}\nonumber\\
S_{GU} & = & {\displaystyle \int} dxdt \left( - 
\frac{1}{4}F_{\mu\nu}F^{\mu\nu} + \frac{e^2\alpha}{2}A_{\mu}A^{\mu} + 
e(\eta^{\mu\nu} - \epsilon^{\mu\nu})(\partial_{\mu}\phi)A_{\nu} + 
\frac{1}{2}(\partial_{\mu}\phi)^2  \right.\\
&& \hspace{0.8in} \left.  + \frac{\displaystyle (\partial_{\mu}\pi_0)^2}{
\displaystyle 2e^2(\alpha - 1)} + \frac{\displaystyle \pi_0}{\displaystyle 
\alpha-1}[(\alpha-1)\eta^{\mu\nu} + \epsilon^{\mu\nu}](\partial_{\mu}
A_{\nu})\right).\nonumber
\end{eqnarray}
On making the replacement $ \pi_0 = - e\;(\alpha-1)\;\theta $ in (29), we get the {\em same} path 
integral and action as in the Batalin-Fradkin case (27). Here this is 
achieved without introducing extra fields. The extra field $\theta$ of the 
BF method is found here {\em within} the original phase space. Further 
the Wess Zumino terms are the same in both cases. It may also be noted 
that, on comparing the gauge invariant Hamiltonians in (21) and (23), 
the extra terms in the $ \pi_{\theta} $ in (21) have been integrated 
away, and so these do not appear in (27).

\subsection{{O(N) Invariant Nonlinear Sigma Model}}

In the earlier example, gauge invariant observables like the Hamiltonian 
had finite number of terms, either in the new variables (BF method) or 
in the discarded constraint of the GU method. The general formalisms of 
Section 2 showed that these observables in general have infinite number 
of terms. The nonlinear sigma model \cite{nsigma} presents an example 
where the gauge invariant observables have infinite number of terms. 
But these can be rewritten in closed form. Even here the two methods 
give the same results.

The model consists of a multiplet of $N$ real scalar fields $n^a, ~a = 
1,2,\ldots N$ and is described by the Lagrangian density
\begin{equation}
{\cal L} = \frac{1}{4}(\partial_{\mu}n^a)(\partial^{\mu}n^a) - \lambda
(n^an_a - 1),
\end{equation}
where $ \lambda$ is a Lagrange multiplier. The canonical Hamiltonian 
density is
\begin{equation}
{\cal H}_c = \pi^a\pi_a + \frac{1}{4}(\partial_1n^a)(\partial_1n_a) + 
\lambda(n^an_a - 1),
\end{equation}
with $ \pi_a = \frac{\displaystyle {\dot n_a}}{2},$ the conjugate momenta.
 The constraints are of the second class,
\begin{equation}
Q_1 = (n^an_a - 1) \approx 0, \hspace{1in} Q_2 = n^a\pi_a \approx 0,
\end{equation}
\begin{equation}
\hspace{0.4in} \{ Q_1(x), Q_2(y) \} = 2|n|^2 \delta(x-y) = 2(Q_1 + 1) 
\approx 2.
\end{equation}
\vspace{1mm}
\noindent The form of $ \lambda $ can be fixed by demanding time 
independence of $ Q_1, Q_2.$  We then get 
\begin{equation}
H_T = {\displaystyle \int} dx \left[ \pi_a\pi_a {|n|^2} + 
\frac{n^a\partial_1^2{n_a}}{4} ~(|n|^2 - 2)\right ].
\end{equation}  
This total Hamiltonian ensures time independence of the constraints 
(32) on the constrained surface defined by {\em both} these constraints.

\vspace{4mm}
We first apply the Gauge Unfixing method. Using (33) we rescale $Q_2$ and 
rewrite as
\begin{equation}
\chi = - \frac{Q_2}{2|n|^2}, \hspace{1.1in} \psi = Q_1 = |n|^2 - 1.
\end{equation}
Choosing $ \chi \cong 0 $ as our first class constraint, we disregard 
$\psi = 0 $
as a constraint. Since the original Hamiltonian $ H_T $ is not invariant 
with respect to $ \chi $ on the new surface defined by $ \chi \cong 0 $, 
we construct and use a projection operator of the form (14). Here we do 
not apply this directly on $ H_T $ to get the gauge invariant Hamiltonian; 
instead we first apply the operator on the fields $ n^a, \pi_a $ to get 
their {\em gauge invariant} analogs. We find that an infinite series of 
the form (14) is required here. For the $ n^a, \pi_a $, these series can 
be rewritten in closed form. The results are, 
\begin{equation}
{\widetilde n}^a_{_{(GU)}} = n^a \left (1 - \frac{\psi}{|n|^2}\right)^{1/2} 
\hspace{8mm} {\sf and} \hspace{8mm} {\widetilde \pi}_{a_{(GU)}} = (\pi_a + 
2n_a\chi)\left(1 - \frac{\psi}{|n|^2}\right)^{- 1/2}.
\end{equation}
These satisfy $ \{ \chi, {\widetilde n}^a_{_{(GU)}} \} = 0 $ and 
$ \{ \chi, {\widetilde \pi}_{a_{(GU)}} \} = 0 $.
Using a property \cite{RaVy} of such projected fields, we substitute 
these gauge invariant fields in $H_T$, and get our gauge invariant 
Hamiltonian,
\begin{eqnarray}
{\widetilde H}_{T_{(GU)}} & = & {\displaystyle \int} dx \left[ (\pi_a + 
2n_a\chi)^2 |n|^2 + \frac{\displaystyle {\widetilde n}^a_{_{(GU)}}
\partial^2_1{\widetilde n}_{a_{(GU)}}}{4}\left(|n|^2 - \psi 
- 2\right)\right]\nonumber\\
& = & {\displaystyle \int} dx \left[ (\pi_a + 2n_a\chi)^2 |n|^2 -
\frac{\displaystyle {\widetilde n}^a_{_{(GU)}}\partial^2_1{\widetilde n}_{
a_{(GU)}}}{4}
\right],
\end{eqnarray}
where we have used (35). It can be verified that ${\widetilde H}_{T_{(GU)}}$
satisfies $ \{\chi, {\widetilde H}_{T_{(GU)}} \} = 0$. This $ {\widetilde 
H}_{T_{(GU)}} $ together with the $ \chi = 0 $ describes a gauge theory here.

\vspace{4mm}
We now apply the Batalin - Fradkin method to this model. We first make 
the choice
\begin{eqnarray}
\omega = 2 \left( \begin{array}{c}
	 0 ~~~ 1\\
	-1 ~~~ 0
\end{array}                   \right) \delta(x-y) \hspace{0.8in}
X (x,y) = \left( \begin{array}{c}
				1 ~~~~~~~~ 0\\
				0 ~~~~ -|n|^2
\end{array}		 \right) \delta(x-y)
\end{eqnarray}
so that the new (first class) constraints are 
\begin{equation}
{\widetilde Q}_1(x) = Q_1 + \Phi^1 \approx 0  \hspace{0.5in}{\sf 
and}\hspace{0.9in} {\widetilde Q}_2(x) = Q_2 - |n|^2\Phi^2 \approx 0,
\end{equation} 
with $ \Phi^1$ and $ \Phi^2$ being the new variables introduced to 
enlarge the phase space (they are not exact canonical conjugates).

With respect to these first class constraints, the gauge invariant 
Hamiltonian is obtained by resorting to the general series (10). Even here
we do not directly construct this Hamiltonian; we look for gauge invariant 
analogs of the $n^a, \pi_a$. Using an infinite series of the form (10) we 
get closed form expressions,
\begin{equation}
{\widetilde n}_{_{(BF)}}^a = n^a\left(1 + \frac{\Phi^1}{|n|^2}
\right)^{\frac{1}{2}} \hspace{1in} {\widetilde \pi}_{a_{(BF)}} = 
(\pi_a - n_a\Phi^2)\left(1 + \frac{\Phi^1}{|n|^2}\right)^{-\frac{1}{2}}.
\end{equation}
Replacing the $ n^a $ and the $ \pi_a $ in $ H_T $ by the $ {\widetilde 
n}^a_{_{(BF)}} $ and the $ {\widetilde \pi}_{a_{(BF)}} $ we get the 
gauge invariant Hamiltonian 
\begin{eqnarray}
{\widetilde H}_{T_{(BF)}} & = & {\displaystyle \int} dx \left[|{
\widetilde \pi}_{_{(BF)}}|^2 |{\widetilde n}_{_{BF}}|^2 + \frac{{ 
\widetilde n}^a_{_{BF}}
\partial_1^2 {\widetilde n}_{a_{(BF)}}}{4} (|{\widetilde n_{_{BF}}}|^2 
- 2)\right]\nonumber\\
& = & \rule{0mm}{8mm}{\displaystyle \int} dx \left [(\pi_a - n_a\Phi^2)^2 
|n|^2 - \frac{{\widetilde n}^a_{_{BF}}\partial_1^2{\widetilde n}_{a_{
(BF)}}}{4}({\widetilde Q}_1 - 1)\right].
\end{eqnarray}
which maintains  the time consistency of the two first class 
constraints in (39). These constraints together with the $ {\widetilde 
H}_{T_{(BF)}} $ describe a gauge theory in the BF method.

\vspace{1mm}
On comparing the gauge invariant observables in (36) of the GU method
and the gauge invariant observables in (40) of the BF method, we 
see that they are the same if we make the identification $ \Phi^1 
= - \psi $ and $ \Phi^2 = - 2\chi $. Obviously due to this identification,
the gauge invariant Hamiltonians in (37) and (41) are also the same, 
apart from the term in $ {\widetilde Q}_1 $ in (41). Thus even here 
extra variables are not required to get gauge symmetries. What comes 
out as an extra variable in the BF method can actually be found in the 
original phase space in the GU method.

\vspace{4mm}
We look at path integral quantisation for the gauge theories obtained in 
these two methods. For the $ {\widetilde H_{T_{(GU)}}}$, the partition 
function is 
\begin{equation}
{\cal Z}_{GU} = {\displaystyle \int} {\cal D}(n^{a}, \pi_{a}, \mu) 
~~{\exp {\left(i{\displaystyle \int} dxdt \left [\pi_a{\dot n^a} - {\cal 
{\widetilde H}}_{T_{(GU)}} - \mu\chi \right]\right)}},
\end{equation}
with $ \mu $ being an arbitrary Lagrange multiplier. We make the 
transformations 
$$
\mu \rightarrow \mu^{\;\prime} = \left(\frac{\displaystyle \mu}{
\displaystyle 2|n|^2} + (n^a\pi_a)\right) \hspace{0.4in}{\sf and} 
\hspace{0.4in} \pi_a \rightarrow \pi_a^{\;\prime} = \left(\pi_a  - 
\frac{\displaystyle{\dot n}_a}{\displaystyle 2|n|^2} - \frac{
\displaystyle \mu^{\;\prime} n_a}{\displaystyle 2|n|^2} \right)
$$ 
and then integrate over the $ \pi^{\;\prime} $. We get 
\begin{equation}
{\cal Z}_{GU} = {\displaystyle \int} {\cal D}(n^{a}, \mu^{\;\prime}) 
~({\sf det}|-n^an_a|)^{1/2} ~{\exp {\left(i{\displaystyle \int} dxdt 
\left [ - \frac{\displaystyle (\partial_1{\widetilde n}^a_{_{GU}}) 
(\partial_1{\widetilde n}_{a_{(GU)}})}{\displaystyle 4} + |n|^2(\mu^{
\prime\prime})^2
\right]\right)}},
\end{equation}
where the $ \mu^{\;\prime\prime} $ is the (once again) redefined 
arbitrary multiplier $ \mu^{\;\prime\prime} = \left(\frac{\displaystyle 
\mu^{\prime}n^a}{\displaystyle 2|n|^2} + \frac{\displaystyle{\dot n}^a}{
\displaystyle 2|n|^2}\right). $ It may be noted from (36) that $ \psi $
is contained within the $ {\widetilde n}_{_{GU}}^a $. 

\vspace{4mm}
For the Hamiltonian $ {\widetilde H}_{T_{(BF)}} $, the partition 
function is 
\begin{eqnarray}
{\cal Z}_{BF} & = & {\displaystyle \int}{\cal D}(\pi_a, n^a, \Phi^1, 
\Phi^2, \lambda_1, \lambda_2)~e^{iS_{BF}} \\
S_{BF} & = & {\displaystyle \int} dxdt \left [\pi_a{\dot n^a} + 
 \frac{1}{2}\Phi^2{\dot \Phi^1} - {\widetilde{\cal H}}_{T_{(BF)}} - 
\lambda_1{\widetilde Q_1} - \lambda{\widetilde Q_2} \right],
\nonumber
\end{eqnarray}
with $ \lambda_1, \lambda_2 $ being undetermined Lagrange multipliers. 
We make the transformations 
$$ 
\begin{array}{rcl}
\pi_a \rightarrow \pi_a^{\;\prime} & = & \left(\pi_a - n_a\Phi^2 + \frac{
\displaystyle \lambda_2 n^a}{\displaystyle 2|n|^2} - \frac{\displaystyle 
{\dot n}^a}{\displaystyle 2|n|^2}\right)\\
\lambda_1 \rightarrow \lambda_1^{\prime} & = & \rule{0mm}{8mm}\lambda_1 
+ \frac{\displaystyle {\widetilde n}^a_{_{BF}}\partial_1^2{\widetilde 
n}_{a_{(BF)}}}{\displaystyle 4} \hspace{0.6in} \lambda_2 \rightarrow 
\lambda_2^{\prime} = \left (\frac{\displaystyle \lambda_2n^a}{
\displaystyle 2|n|^2} - \frac{\displaystyle {\dot n}^a}{\displaystyle 
2|n|^2}\right )
\end{array}
$$
and integrate over $ \pi_a^{\;\prime}, ~\lambda^{\prime}_1 $. The 
latter integration gives a delta function $ \delta(\Phi^1 + |n|^2 -1). $
Integration over the $ \Phi^1 $ will replace $ \Phi^1 $ everywhere
by $ - |n|^2 + 1 = - \psi $. We then get
\begin{equation}
{\cal Z}_{BF} = {\displaystyle \int} {\cal D}(n^{a}, 
\lambda^{\prime}_2) ~({\sf det}|-n^2|)^{1/2}~{\sf exp} \left({\displaystyle 
i \int {\sf dxdt} \left[ |n|^2(\lambda^{\prime}_2)^2 - \frac{\displaystyle 
(\partial_1{\widetilde n}^a_{_{BF}})(\partial_1{\widetilde n}_{a_{(BF)}})} 
{\displaystyle 4}\right]}\right).
\end{equation}
Due to the delta function $ \delta(\Phi^1 + |n|^2 -1) $, the $ \Phi^1 $
in the expression (40) for $ {\widetilde n}^a_{_{BF}} $ is now replaced 
by $ - \psi = (-|n|^2 + 1), $ so that from (36) we now have $ {\widetilde 
n}^a_{_{BF}} = {\widetilde n}^a_{_{GU}} $. Using this, and taking note 
of the arbitrary nature of the multipliers $ \mu^{\prime\prime} $ in 
(43) and the $ \lambda^{\prime} $ in (45), we see that we get the {\em 
same results} from both the BF and the GU methods!

\section{{General Proof for Two Second Class Constraints}}

Having considered the examples in the earlier section, we now arrive 
at a general proof of the equivalence between the Batalin-Fradkin and 
Gauge Unfixing methods. We consider the case of two second class 
constraints. 

In the ~{\sf gauge unfixing} method, we redefine the two constraints as 
\begin{equation}
\chi = \frac{1}{\sf E}Q_1, \hspace{1in} \psi = Q_2,
\end{equation}
where $ {\sf E}(q,p) = \{Q_1, Q_2\}. $ We retain the $ \chi $ as the 
first class constraint, and discard the 
$ \psi$ (other choices are also possible). The construction and 
application of the corresponding {\sf projection operator} $ I\!\!P $ 
on a phase space function ${\sf A}$ gives the gauge invariant function
$$
{\widetilde A}_{GU} = {\displaystyle :\;e^{\displaystyle 
-\psi{\hat \chi}}:}\;{\sf A} ~= ~{\sf A} - \psi\{\chi, {\sf A}\} + 
\frac{\psi^2}{2!}\{\chi,\{\chi, {\sf A}\}\} - \frac{\psi^3}{3!} \{\chi,
\{\chi,\{\chi,{\sf A}\}\}\} + \ldots 
\eqno(14)
$$
with an infinite number of terms. Of these, apart from the $ {\sf A} $, 
we give below terms upto the fourth order. Using (46) and $ {\sf E} 
= \{ Q_1, Q_2 \} $, these terms are 
\begin{eqnarray}
- Q_2 \{ \chi, {\sf A} \} & = & \rule{0mm}{8mm}- \frac{Q_2}{\sf E}\{Q_1, 
{\sf A} \} + \ldots\ldots\ldots\ldots \nonumber\\
+ \frac{Q_2^2}{2!} \{ \chi, \{\chi, {\sf A}\}\} & = & \rule{0mm}{10mm}
\frac{1}{2!}\;\frac{Q_2^2}{\sf E^2}\left[\{Q_1, \{Q_1, {\sf A} \}\} + {\sf 
E}\left\{Q_1, \frac{1}{\sf E}\right\}\{Q_1,{\sf A}\}\right] + 
\ldots\ldots\ldots\ldots\\ 
- \frac{Q_2^3}{3!} \{\chi,\{\chi,\{\chi, {\sf A}\}\}\} & = & 
\rule{0mm}{10mm}- \frac{1}{3!}\frac{Q_2^3}{\sf E^3}\left[\{Q_1, \{Q_1,
\{Q_1,{\sf A} \}\}\} + 3{\sf E}\left\{Q_1,\frac{1}{\sf E}\right\} \{Q_1,
\{Q_1, {\sf A}\}\} \right.\nonumber\\
&&  ~~~~\rule{0mm}{8mm}\left.+ ~{\sf E^2} \left\{Q_1,\frac{1}{\sf E}
\right\}^2\{Q_1,{\sf A}\} + {\sf E}\left \{Q_1,\left\{Q_1,\frac{1}{\sf 
E}\right\}\right\} \{Q_1,{\sf A}\}\right] \nonumber\\
&& \hspace{0.5in} + \ldots\ldots\ldots\ldots\ldots\ldots\ldots\ldots
\nonumber
\end{eqnarray}
\begin{eqnarray}
+ \frac{Q_2^4}{4!} \{\chi,\{\chi,\{\chi,\{\chi, {\sf A}\}\}\}\} & = & 
\rule{0mm}{10mm}
\frac{1}{4!}\frac{Q_2^4}{\sf E^4} \left(\{Q_1,\{ Q_1,
\{Q_1\{Q_1, {\sf A}\}\}\}\} + 6\;{\sf E}\left\{Q_1, \frac{1}{\sf E}
\right\}\{Q_1,\{Q_1,\{Q_1,{\sf A}\}\}\} \right.\nonumber\\
&&\rule{0mm}{8mm} \hspace{5mm}\left.+ \;7\;{\sf E^2}\left\{Q_1,
\frac{1}{\sf E}\right \}^2 \{Q_1,\{Q_1,{\sf A}\}\} + {\sf E^3}\left\{Q_1,
\frac{1}{\sf E}\right\}^3 \{Q_1, {\sf A} \} \right.\nonumber\\
&& \rule{0mm}{8mm} \hspace{5mm}\left.+ \;4\;{\sf E}\;\left\{Q_1\left\{Q_1,
\frac{1}{\sf E}\right\}\right\} \;\left[\{Q_1,\{Q_1,{\sf A}\}\} + {\sf 
E}\left\{Q_1, \frac{1}{\sf E}\right\}\{Q_1, {\sf A}\}\right]\right.   
\nonumber\\
&& \rule{0mm}{8mm} \hspace{5mm}\left.+ \;{\sf E}\;\left\{Q_1,\left\{Q_1,
\left\{ Q_1,\frac{1}{\sf E}\right\}\right\}\right\} \{Q_1, {\sf A} \} 
\right) \;+ \;\ldots\ldots\ldots\ldots\ldots \nonumber
\end{eqnarray}
In the right hand sides of each equation in (47) above we have explicitly 
given only those terms which are proportional to only the $ \psi \;(= Q_2)$. 
There are other terms, which are {\sf proportional} to the first class 
constraint $ \chi$. These terms can however be put to zero, by using 
$ \chi = 0 $.

\vspace{4mm}
In the BF method, the (modified) first class constraints have the general 
form (5), an infinite series in the new variables. We will consider the 
case where this series is truncated after the second term. Since the 
choice of the matrices $ X_{ab} $ and $ \omega^{ab} $ of eqn. (7)
 reflects the arbitrariness in the new gauge theory, we make here a 
specific choice, 
\begin{eqnarray}
\!\!\!\!\!\!\!\!\!  \omega^{ab} = \left( \begin{array}{c}
                0\;\;\;\;1\\
                 \!\!-1\:\;\;\;0
                  \end{array}\right ), ~~~~~~~~~
                      & ~~~~~~X_{ab} = \left( \begin{array}{c}
                                        X\;\;\;0\\
                                         0\;\;\;\;1
                                          \end{array}\right )\\
&&\nonumber\\\hspace{19mm} X  = - \;{\sf E}, & {}
\end{eqnarray}
with $ {\sf E}(q,p) = \{ Q_1, Q_2 \} $. Here (49) is obtained by 
substituting (48) in the first order equation (7). Using (48), 
(49) and (9), and by equating the second order term from (9) to 
zero, we get the condition for truncating the series (5) after 
the second term as
\begin{equation}
\{ Q_2, {\sf E} \} = 0.
\end{equation} 
Higher order terms are also zero. We will assume (50) from now on. We 
thus get the new first class constraints,
\begin{equation} 
{\widetilde Q}_1 = Q_1 - {\sf E}\;\Phi^1, \hspace{1.4in} 
{\widetilde Q}_2 = Q_2 + \Phi^2, 
\end{equation}
For the choice (48) we now look at the various terms in the series (10) 
for a general gauge invariant variable. If for example we consider the 
first and second order terms,
\begin{eqnarray}
{\sf A}^{(1)} & = & ({\widetilde Q}_a - Q_a)({\cal E}^{-1})^{ab}
\{Q_b, {\sf A} \},\nonumber\\
{\sf A}^{(2)} & = & \rule{0mm}{8mm}({\widetilde Q}_a - Q_a)({\cal 
E}^{-1})^{ab}\frac{1}{2} ({\widetilde Q}_c - Q_c)({\cal E}^{-1})^{cd}
\left[\{Q_b,\{Q_d, A \}\} + X_{de}\{Q_b, X^{ef}\}\{Q_f, A\}\right]\nonumber\\
&& \hspace{1.4in}+ \frac{1}{2}\frac{({\widetilde Q}_2 - Q_2)}{\sf E}
\frac{({\widetilde Q}_1 - Q_1)}{\sf E}\{{\sf E}, A \},\nonumber
\end{eqnarray}
we see that terms proportional to both $ ({\widetilde Q}_1 - 
Q_1) $ and $ ({\widetilde Q}_2 - Q_2) $ are present. The higher 
order terms will also have terms proportional to the $ ({\widetilde 
Q}_1 - Q_1) $ and $ ({\widetilde Q}_2 - Q_2) $. Since both $ 
{\widetilde Q}_1 $ and $ {\widetilde Q}_2 $ are first class constraints 
in the BF construction, we can ignore the terms proportional to $ 
{\widetilde Q}_1 $ and $ {\widetilde Q}_2 $. We are then left with terms 
separately proportional to the $ Q_1 $ and $ Q_2 $, and terms 
containing the product $ Q_1Q_2$. We then have, upto the fourth order, 
\begin{eqnarray}
A^{(1)} & = & - \;\frac{Q_2}{\sf E} \{Q_1, A\} + ~\ldots\ldots\nonumber\\
A^{(2)} & = & \rule{0mm}{8mm}+ \frac{1}{2!}\left(\frac{Q_2}{\sf E}\right)^2\left[\{Q_1,
\{Q_1,A\}\} + {\sf E}\left\{Q_1, \frac{1}{\sf E}\right\}\{Q_1,A\}\right] 
+ ~\ldots\ldots\nonumber\\
A^{(3)} & = & \rule{0mm}{8mm} -\; \frac{1}{3!}\left(\frac{Q_2}{\sf E}
\right)^3\left[\{Q_1,\{Q_1,\{Q_1,A\}\}\} + 3{\sf E}\left\{Q_1, 
\frac{1}{\sf E}\right\}\{Q_1, \{Q_1,A\}\} \right.\\
&&\left. \hspace{0.8in} + ~{\sf E} \left\{Q_1,\left\{Q_1,\frac{1}{\sf E}
\right\}\right\}\{Q_1, A\} + {\sf E}^2\left\{Q_1,\frac{1}{\sf E}\right
\}^2\{Q_1,A\}\right]\nonumber\\
&& \hspace{1in}+ \ldots\ldots\ldots\ldots\ldots\ldots\ldots\nonumber
\end{eqnarray}
\begin{eqnarray}
A^{(4)} & = & \rule{0mm}{10mm}\frac{1}{4!}\left(\frac{Q_2}{\sf E} \right)^4 
\left[ ~\{Q_1,\{Q_1,\{Q_1,\{Q_1,A\}\}\}\} + 6\;{\sf E}\left\{Q_1,
\frac{1}{\sf E}\right\} \{Q_1,\{Q_1,\{Q_1, A\}\}\} \right.\nonumber\\
&&\rule{0mm}{8mm}\hspace{18mm}\left. + \;7 \;{\sf E}^2\left\{Q_1,
\frac{1}{\sf E}\right\}^2 \{Q_1,\{Q_1, A\}\} + {\sf E}^3\left\{Q_1,
\frac{1}{\sf E}\right\}^3\{Q_1, A\} \right.\nonumber\\ 
&&\rule{0mm}{8mm}\left. \hspace{2cm}+ \;4 {\sf E} \;\left\{Q_1,\left\{
Q_1,\frac{1}{\sf E} \right\} \right\}\left(\{Q_1,\{Q_1, A\}\} \;+ \;{\sf 
E}\left \{Q_1,\frac{1}{\sf E}\right\}\{Q_1, A\}\right)\right.\nonumber\\
&&\rule{0mm}{8mm}\hspace{2cm}\left. + \;{\sf E}\left\{Q_1,\left\{Q_1,
\left\{Q_1,\frac{1}{\sf E}\right\}\right\}\right\}\{Q_1, A\} \right] 
+ ~\ldots\ldots\ldots\ldots\ldots\ldots\nonumber
\end{eqnarray}
where we have explicitly given terms proportional to only the $ Q_2 $. 
The terms proportional to the $ Q_1$ will be, as explained below, ignored.

To compare the gauge invariant ${\widetilde A}_{BF}$ and ${\widetilde 
A}_{GU}$ of the two methods, we look at the terms of different orders 
in (47) and (52). It can be seen that, for each of the orders considered, 
the terms proportional to the $ \psi \;(= Q_2)$ in (47) are {\em the same} 
as those proportional to the $ Q_2 $ in (52). Even though this is shown 
here for terms upto the fourth order, it can be verified to be true for 
higher terms also. Both (47) and (52) will also have terms proportional 
to the $ Q_1 $ (or $ \chi $), though these need not be the same. We thus
conclude that
\begin{equation}
{\widetilde A}_{BF} = {\widetilde A}_{GU} + \sum_{m=1}^{\infty} (~~)Q_1^m.
\end{equation}
Since the second term (a series) on the RHS is proportional to the first 
class constraint $\chi$, it goes to zero on the constraint surface, and so 
it can be ignored. Thus the gauge 
invariant observables from the two methods are {\sf equivalent}
upto terms involving the first class constraint of the {\sf gauge 
unfixing} method. 

\section{{Conclusions}}

We conclude by going back to the question posed in the title of this
paper : Are new variables necessary to extract hidden symmetries in 
second class constrained systems? ~We find that, for a fairly general
class of systems, this question is answered in the negative. Extra 
variables are {\em not necessary}; rather the hidden gauge symmetry 
{\em can be found within} the original system itself.

The above conclusion has been demonstrated by first looking at two 
theories as examples and then by presenting a proof for a fairly 
general second class constrained system. We have shown that in all 
these, the Batalin-Fradkin and Gauge Unfixing  methods, even though 
widely different in construction, give the same results. A similar 
conclusion was also made in an earlier paper \cite{Asvva}.  

In the past gauge invariances have been induced in some systems (like 
anomalous gauge theories) by sometimes introducing what are known as ~{\sf 
compensating fields}. These correspond to the gauge degrees of freedom.
The extra variables of the BF method can be identified
with these compensating fields. Using the GU method we can then say that 
these compensating fields can be found within the original phase space.

The general proof given in Section 4 considers a case where the first
class constraints in the BFT method have a particular form, with terms 
beyond the first order in the extra variables being zero. It is to be 
seen if a similar proof holds for a more general form of the first 
class constraints.

The general proof of Section 4 involved only two second class constraints.
One has to see if a similar proof of equivalence can be obtained for more than
two second class constraints. In this context, before looking for gauge
invariant observables, one must see whether first class constraints can 
always be obtained in both methods. It may be recalled that mention was
made of the $X$ matrix of the BFT method \cite{BaFa} being a ``{\sf 
square root}'' of the $ {\cal E} $ matrix of (2). For more than two 
second class constraints, getting this $X$ matrix may become a non-trivial 
issue in the general case. Similarly, in the GU method, the classification 
into first class and gauge fixing-like constraints in a global manner may 
be a nonitrivial issue \cite{RaVy}. Looking for equivalence  of the two 
methods is to be done only after these issues are resolved.

\vspace{7mm} 
\noindent{\large{\bf Acknowledgements}}

\vspace{4mm}
We thank the Chairman, Physics Department, BU and Prof B A Kagali, BU 
for constant encouragement.

We wish to acknowledge the  ~{\sf Council for Scientific and Industrial 
Research, New Delhi, India} ~for a previous Research Associateship and 
the present Senior Research Associateship (Pool Scheme) of the 
Government of India.

\end{document}